\documentclass[preprint,review,12pt]{elsarticle}

\usepackage{lineno,hyperref}
\usepackage{amssymb,rotate,color,amsmath}
\usepackage{setspace}
\usepackage{graphicx,epsfig,psfrag}
\usepackage{times}

\modulolinenumbers[1]

\journal{Annals of Physics}
\newcommand{\be}{\begin{equation}}
\newcommand{\ee}{\end{equation}}
\newcommand{\bearr}{\begin{eqnarray}}
\newcommand{\eearr}{\end{eqnarray}}
\newcommand{\nn}{\nonumber}

\newcommand{\eps}{\varepsilon}

\newcommand{\bk}{{\bf k}}

\newcommand{\up}{\uparrow}
\newcommand{\down}{\downarrow}

\begin{document}

\begin{frontmatter}

\title{Quantum transport through 3D Dirac materials}

\author[mymainaddress]{M. Salehi}
\author[mymainaddress,mysecondaryaddress]{S. A. Jafari \corref{mycorrespondingauthor}}
\cortext[mycorrespondingauthor]{Corresponding author, Tel:+982166164524.}
\ead{jafari@physics.sharif.edu}

\address[mymainaddress]{Department of Physics, Sharif University of Technology, Tehran 11155-9161, Iran}
\address[mysecondaryaddress]{Center of excellence for Complex Systems and Condensed Matter (CSCM), Sharif University of Technology, Tehran 1458889694, Iran}

\begin{abstract}
Bismuth and its alloys provide a paradigm to realize three dimensional materials whose
low-energy effective theory is given by Dirac equation in 3+1 dimensions. 
We study the quantum transport properties of three dimensional Dirac materials 
within the framework of Landauer-B\"uttiker formalism. Charge carriers in normal 
metal satisfying the Schr\"odinger equation, can be split into four-component
with appropriate matching conditions at the boundary with the three dimensional 
Dirac material (3DDM). We calculate the conductance and the Fano factor of an interface
separating 3DDM from a normal metal, as well as the conductance through a slab of
3DDM. Under certain circumstances the 3DDM appears transparent to electrons hitting
the 3DDM. We find that electrons hitting the metal-3DDM interface from metallic side can
enter 3DDM in a reversed spin state as soon as their angle of incidence deviates
from the the direction perpendicular to interface. However the presence of 
a second interface completely cancels this effect.
\end{abstract}


\end{frontmatter}


\section{Introduction}
\label{Intro}
After discovery of graphene~\cite{novoselov1}, the concept of Dirac fermions became 
a live and daily-life concept to condensed matter physicists. In the regime of low-energy 
excitations, the single-particle excitations in graphene obey an effective Hamiltonian that is
identical to two dimensional Dirac equation~\cite{novoselovnobel}. Some of the intriguing
properties inherited from the relativistic-like form of the underlying Dirac equation are, 
Klein tunneling~\cite{Katsnelson2006}, unconventional Hall effect~\cite{Zhang2005,Novoselov2006}, 
bipolar super-current~\cite{Heersche2007} and so on. 
Parallel to the developments in graphene physics, inspired by original proposal of 
Haldane~\cite{Haldane88} based on the honeycomb lattice structure of graphene, 
Kane and Mele constructed a model for two-dimensional topological insulator (TI)~\cite{Kane2005}.
Later on other models of TIs carrying edge modes due to their non-trivial topology were 
theoretically constructed~\cite{Bernevig2006} and experimentally verified~\cite{Konig}.
Three dimensional counterparts of the TIs displaying gap in the bulk, and massless Dirac
fermions on their surface~\cite{fu2007PRL} were all based on the Bismuth element.

The elemental Bismuth was studied since a long time ago and the low-energy effective theory
around the L point of Brillouin zone was proposed by Wolf~\cite{Wolff1964} based on 
two-band approximation of Cohen~\cite{Cohen1960}. It was found that effective theory 
describing the spin-orbit coupled bands of Bismuth is indeed a three dimensional (3D) massive Dirac 
theory. Later experiments corroborated the picture of 3D Dirac fermions in this 
material~\cite{Li2008}. More recently, massless 3D Dirac fermions were observed at 
the $\Gamma$ point of Brillouin zone of the Na$_3$Bi compound~\cite{Liu2014}. 
This provides us with condensed matter realization of both massive and massless 
Dirac fermions in three spatial dimensions. Therefore it is timely to study the 
transport properties of 3D Dirac electrons in various settings.

In this paper, we investigate the ballistic transport of 3D Dirac fermions across a
boundary separating the 3D Dirac material (3DDM) from the normal metal, as well as
the quantum transport through a segment of 3DDM sandwiched between two metallic
leads as depicted in Fig.~\ref{fig:structure}. 
The dynamics of charge carriers inside the 3DDM is described by the
3D Dirac equation, while the electronic states inside the normal metallic leads
are governed by the scalar Schr\"odinger equation. 
Due to such a difference in the governing equations in the two sides
of the interface, the boundary condition matching the electronic states will be
tricky and one has to choose the wave-functions so as to give identical current
density in both sides of the interfaces separating 3DDM and the normal metals. 
In the following sections we will formulate this problem and will calculate the
transport properties in the ballistic regime within the Landauer-B\"uttiker
formalism.

\section{Formulation of the problem}
\label{theory}
\begin{figure}[bt]
\begin{center}
\includegraphics[width=8cm]{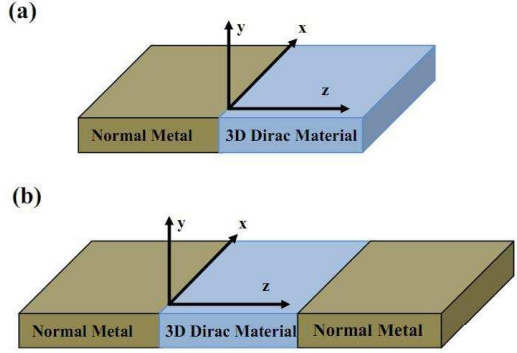}
\caption{\label{fig:structure} (Color online) The structures of junctions between 3D Dirac material 
(3DDM) and normal metals}
\end{center}
\end{figure}

The structure of junctions that we consider in this work is depicted in Fig.~\ref{fig:structure}. 
In metallic region the carriers obey the Schr\"odinger equation that means the wave functions $\phi$ 
is a one-component function; whereas in 3DDM the wave-function $\psi$ describing the charge 
carrier is an four component spinor satisfying the 3D Dirac equation.
An important question is how to construct a boundary condition 
for matching these two different types of wave-functions across the boundary?

The explicit form of the effective Hamiltonian for the 3DDM 
is given by~\cite{Fuseya2009}:
\begin{eqnarray}
H=
\left(
\begin{array}{cccc}
  \Delta                    & 0                        &                 i v_D q_z                  &  i v_D (q_x - i q_y)   \\
  0                            & \Delta                & i v_D (q_x + i q_y)                   & - i v_D q_z                \\
-i v_D q_z                & i v_D (-q_x+ i q_y) & - \Delta                                & 0                               \\
-iv_D (q_x+i q_y)    & i v_D q_z              & 0                                            & -\Delta                      
\end{array}
\right),
\label{wolfhamiltonian}
\end{eqnarray}
where $\Delta$ is energy gap, $v_D$ is velocity of carriers 
and $q_i,{i=x,y,z}$ denotes the Cartesian components of the wave-vector  $\bf q$. 
The above explicit form corresponds to the following choice of $4 \times 4$ 
Dirac matrices:
\begin{eqnarray}
\alpha_i=
\left(
\begin{array}{cc}
  0                    & i \sigma_i        \\
 - i \sigma_i      & 0                      \\
\end{array}
\right),~~~~~~
\beta =
\left(
\begin{array}{cc}
\sigma_0           & 0              \\
0                  &-\sigma_0       \\
\end{array}
\right)
\label{alphabeta}
\end{eqnarray}
where $\sigma_i$ are Pauli matrices and $\sigma_0$ is $2 \times 2 $ 
unit matrix. The $\gamma_0= \beta$  and other $\gamma_i $ matrices  are defined as
\begin{equation} 
\gamma_i =v_D \beta \alpha_i=
 v_D \left(
\begin{array}{cc}
0                    &  i \sigma_i \\
i \sigma_i        & 0             \\
\end{array}
\right).
\label{sigma} 
\end{equation}
With the above choice, Hamiltonian can be compactly written as
\begin{equation}
   H=\Delta \beta + {\bf q}.{\bf \gamma}.
\end{equation}

The $i$'th component of the current density in 3DDM region is given by 
\begin{equation}
j_i= \bar{\psi}\gamma_i \psi
\label{diraccurrent}
\end{equation}
where $\bar{\psi}=\psi^{\dagger} \gamma_0$ and the $\gamma_i$ matrices are given
by Eq.~\eqref{sigma}.

The wave-function $\phi$ in the metallic region
($ x < 0$ in Fig.~\ref{fig:structure}) satisfies the Schr\"odinger equation,
\begin{equation}
   -\frac{\hbar^2}{2m}\nabla^2 \phi=(\xi+\mu_m) \phi\equiv \eps \phi,
\end{equation}
with $m$ being effective mass of metallic carriers,
and $\mu_m$ is the chemical potential in the normal metallic region with respect
to which energy $\xi$ is measured. 
The current density resulting from the above wave-function is
  \begin{equation}
    {\bf j}=\frac{\hbar}{2im}(\phi^* \nabla \phi - \phi \nabla \phi^*).
    \label{schrodingercurrent}
    \end{equation}

\subsection{The basis in the 3DDM and normal metallic regions}
We are going to use the Landauer-B\"uttiker formulation in order to obtain
the transport properties of 3DDM. For this we need to fix a basis with
respect to which all the amplitudes will be built. 

\begin{figure}[tb]
\begin{center}
\includegraphics[width=8 cm]{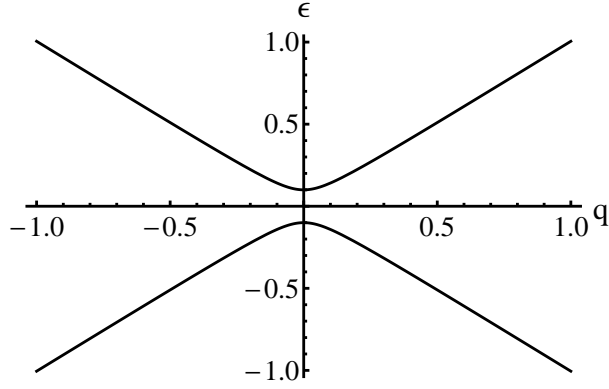}
\caption{\label{fig:dispersion} (Color online) The dispersion relation given by Eq.~(\ref{dispersion}). 
The zero of energy corresponds to the middle of the gap.}
\end{center}
\end{figure}

Let us first discuss the normal metallic region. 
In this region the one-particle wave-equation is $H \phi=(\xi+\mu_m) \phi=\eps \phi$ with
$\xi = \hbar^2 k^2/2m$, where $\mu_m$ has been introduced to allow for 
possible difference in the origin of energies in the metallic and 3DMM sides.
The wave-function in normal metal is,
\begin{equation}
\phi=A e^{i \bf{k.r}}+B e^{-i \bf{k.r}}.
\label{metalwave}
\end{equation}
where $\bk$ is the wavenumber in normal metal region. $A$ and $B$ are the amplitudes of 
right-going and left-going waves, respectively. For this wave function, the current
density, Eq.~\eqref{schrodingercurrent} will be given by
\begin{equation}
j_i=\frac{\hbar k_i}{m}(|A|^2-|B|^2).
\label{metalcurrent}
\end{equation}
To equate this current density to the one arising from solutions of 3D Dirac equation,
we need to obtain the expression for the current density in the 3DDM region.
With the Hamiltonian in Eq.~\eqref{wolfhamiltonian} the wave equation 
$H \psi= \eps \psi$ eigen-energies are,
\begin{equation}
\eps= \pm \sqrt{v_{D}^2 q^2 + \Delta^2}=\eps_{\pm}
\label{dispersion}
\end{equation}
where $\eps_{+ (-)}$ refers to conduction (valence) band hosting electron (hole) excitations. 
In Fig.(\ref{fig:dispersion}), the dispersion relation of Bi is plotted.
Each band has a two-fold spin degeneracy.
For $\eps_{+}$ we have two wave function that corresponds to spin up and down, 
and are given by,
\begin{equation}
\psi_{e \uparrow}=
\left(
\begin{array}{c}
1 \\
0 \\
\frac{-i v_D q_z}{\eps_{+}+\Delta} \\
\frac{-i v_D (q_x+iq_y)}{\eps_{+}+\Delta}\\
\end{array}
\right)
e^{i {\bf q.r}},
\psi_{e \downarrow}=
\left(
\begin{array}{c}
0 \\
1 \\
\frac{-i v_D (q_x-i q_y)}{\eps_{+}+\Delta} \\
\frac{i v_D q_z}{\eps_{+}+\Delta}\\
\end{array}
\right)
e^{i {\bf q.r}}.
\label{biwave}
\end{equation}
For $\eps_{-}$ case, the wave functions of spin up and spin down hole states are:
\begin{equation}
\psi_{h \uparrow}=
\left(
\begin{array}{c}
\frac{i v_D q_z}{\eps_{-}-\Delta} \\
\frac{i v_D (q_x+iq_y)}{\eps_{-}-\Delta} \\
1 \\
0\\
\end{array}
\right)
e^{i {\bf q.r}},
\psi_{h \downarrow}=
\left(
\begin{array}{c}
\frac{i v_D (q_x-i q_y)}{\eps_{-}-\Delta} \\
\frac{-i v_D q_z}{\eps_{-}-\Delta}\\
 0\\
1\\
\end{array}
\right)
e^{i {\bf q.r}}.
\label{biwave2}
\end{equation}
When the electrons moving to the right in the normal region reach the interface between the
3DDM and the normal region, the current density of right movers will be given
by contributions coming from $\up$ and $\down$ spin states. This allows us to 
write $|A|^2=|\alpha_{\uparrow}|^2+|\alpha_{\downarrow}|^2$, where $|\alpha_{\uparrow}|^2$ 
($|\alpha_{\downarrow}|^2$) is the contribution of spin $\up$ ($\down$) electrons that 
travel in positive direction inside the normal metal. Corresponding to amplitude $B$ of the 
left-movers in the normal metal, there are amplitudes $\beta_\up$ and $\beta_\down$ 
of the spin-$\up$ and spin-$\down$ left-movers in metallic region satisfying 
$|B|^2=|\beta_{\uparrow}|^2+|\beta_{\downarrow}|^2$.
With this, Eq.~\eqref{metalwave} in the normal region can be resolved in terms of
its $\up$ and $\down$ components in the form of
\begin{equation}
\phi= 
\left(
\begin{array}{c}
\alpha_{\uparrow} \\
\alpha_{\downarrow}\\
\end{array}
\right)
e^{i \bf{k.r}}+
\left(
\begin{array}{c}
\beta_{\uparrow}\\
\beta_{\downarrow}\\
\end{array}
\right)
e^{-i \bf{k.r}},
\label{metalwave2}
\end{equation}
where up to this point the spinorial notation merely indicates that the wave function satisfies
two copies of Sch\"odinger equation. 
At this point following Sepkhanov and coworkers~\cite{Sepkhanov2007}, 
we construct a {\em virtual four-component} wave function for {\em metal region} as:
\begin{equation}
\Phi=
\left(
\begin{array}{c}
i \alpha_{\uparrow}\\
\alpha_{\downarrow}\\
\alpha_{\uparrow}\\
i \alpha_{\downarrow}\\
\end{array}
\right)
e^{i \textbf{k.r}}+
\left(
\begin{array}{c}
i \beta_{\uparrow}\\
-\beta_{\downarrow}\\
- \beta_{\uparrow}\\
i \beta_{\downarrow}\\
\end{array}
\right)
e^{-i \textbf{k.r}},
\label{virtualwavemetal}
\end{equation}
   The above form has chosen in such a way that when inserted in Eq.~(\ref{diraccurrent}) 
that comes from Dirac equation, 
gives rise to the current given by Eq.~(\ref{metalcurrent}) that is based on the Schr\"odinger
equation. Therefore we now have a four components wave function which can be used in matching 
the wave functions at interface. In case of junctions where the velocity $v_m$ in the metallic side
and $v_D$ in the 3DDM side are different a $1/\sqrt{v_m}$ pre-factor with $v_m=\hbar k / m$ must 
be multiplied to $\Phi$ in Eq.~(\ref{virtualwavemetal}). Similarly a $1/\sqrt{v_D} $ pre-factor 
must be multiplied to Eq.~{\eqref{biwave}} to preserve the unitarity of the S-matrix of quantum 
transport~\cite{Nazarov2009}. These pre-factor will cancel the $v_D$ in definition of $\gamma$ 
matrices for 3DDM side and the $v_m$ in the expression for the current density of the normal 
metallic side. 
The current density in the normal metal side must be equal to the one in the 3DDM side.
With the above point in mind it takes the following form:
   \begin{equation}
   \bar{\psi}\frac{\gamma_z}{v_D}\psi=\bar{\Phi}\frac{\gamma_z}{v_m} \Phi.
   \end{equation}
This condition will be satisfied when the matching condition $\psi=\Phi$ at the interface
is imposed. This matching condition determines transmission and reflectance coefficients out of which the
transport properties can be calculated in standard way. It is important to note that the 
choice in Eq.~(\ref{virtualwavemetal}) for the form of the virtual four-component wave function
in the metallic side is not unique. However as long as measurable quantities such as
transmission or reflectance are concerned, this choice does not matter and any
arbitrary choice will give the same result. Therefore we work with the form given
in Eq.~(\ref{virtualwavemetal}). 

\section{Results and Discussion}
\subsection{Metal-3DDM junction}
In this section we consider the transport of positive energy ($\eps=\eps_+$) states
corresponding to electron-like particles. The transport of hole-like excitations
will be similar. 
 In Fig.~\ref{fig:structure}~(a)  we consider an interface separating a metallic region
from the 3DDM region. An electron coming from $x\rightarrow -\infty$ has two possibilities: 
either passes thorough interface with probability amplitude $t$ or reflects to the left with
amplitude $r$. Assuming that the incident electron has spin $\up$, and putting this in the 
four-component form, Eq.~(\ref{virtualwavemetal}), the 
wave function in the normal region can be written as,
 \begin{equation}
 \Phi=
 \left(
 \begin{array}{c}
 i \\
 0\\
 1\\
 0\\
 \end{array}
 \right)
 e^{i \textbf{k.r}}+
 \left(
 \begin{array}{c}
 i r_{\uparrow}\\
 -r_{\downarrow}\\
 -r_{\uparrow}\\
 i r_{\downarrow}\\
 \end{array}
 \right)
 e^{-i \textbf{k.r}}.
 \label{spinupvirtualwavemetal}
 \end{equation}
where $r_{\up}$ ($r_\down$) corresponds to reflection coefficient of back scattered 
electron with spin $\up$ ($\down$). 
On the other hand the transmitted electron now satisfies the 3D Dirac equation
and hence inside the 3DDM region the wave function is of the form,
 \begin{equation}
 \psi=t_{\uparrow} \psi_{\uparrow}^{+}+t_{\downarrow}\psi_{\downarrow}^{+}
 \label{boundarycondition1}
 \end{equation}
where $t_{\uparrow}$ ($t_{\downarrow}$) is the transmission amplitude for entering the 3DDM
region as an electron with $\up$ ($\down$). Matching the two wave functions in 
Eq.~(\ref{spinupvirtualwavemetal}) and Eq.~(\ref{boundarycondition1}) gives the following 
result for the transmission in the spin $\up$ and $\down$ channels, respectively:
\bearr
&&t_{\uparrow}=
\left(
\frac{(\eps+\Delta)(\eps+ \Delta+v_D q_z)}{(\eps+ \Delta+ v_D q_z)^2+(v_{D}^2(q_{x}^2+q_{y}^2))}
\right),
\label{tup}\\
&&t_{\downarrow}=
\left(
\frac{(\eps+\Delta)(v_D \sqrt{q_{x}^2+q_{y}^2})}{(\eps+\Delta+v_D q_z)^2+(v_{D}^2(q_{x}^2+q_{y}^2))}
\right).
\label{tdown}
\eearr
The above spin-resolved transmission probabilities give the total transmission 
\begin{equation}
T=|t_{\uparrow}|^2+|t_{\downarrow}|^2=
\left(
\frac{(\eps+\Delta)^2}{(\eps+\Delta+v_D q_z)^2+(v_{D}^2(q_{x}^2+q_{y}^2))}
\right).
\label{totalt}
\end{equation}
It is interesting to note that according to the solutions Eq.~\eqref{tup} and~\eqref{tdown},
a spin-$\up$ electron incident from the normal region not only can be transmitted as
spin-$\up$, but can also be transmitted as spin-$\down$ electron. The probability 
amplitude for the later process is given by $t_\down$. This interesting feature
is absent for normal incidence where $q_x=q_y=0$. {\em Therefore electrons hitting the
3DDM at an angle have the chance of being transmitted into 3DDM as spin-flipped}.
There are also two interesting limiting behavior of the above expressions
for the case of normal incidence: (i) In the " ultra-relativistic" situation where
$\eps\gg \Delta$, the total transmission probability always tends to $1/4$. 
(ii) When the energy of the incident particle is such that it is injected to the
bottom of the positive energy states in the 3DDM, the transmission probability
equals $1$. 


\begin{figure}[tb]
\begin{center}
\includegraphics[width=6.2cm]{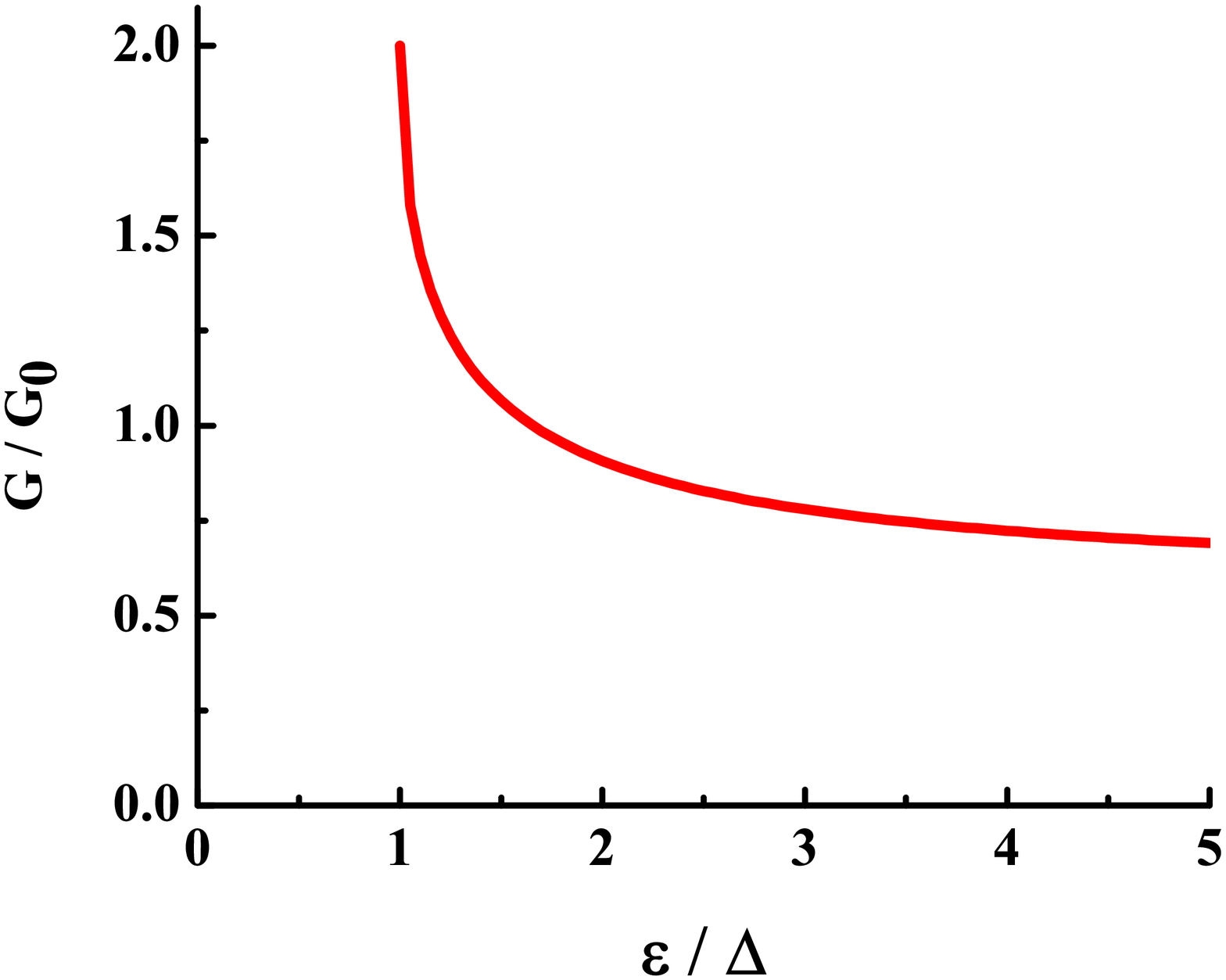}
\includegraphics[width=6.2cm]{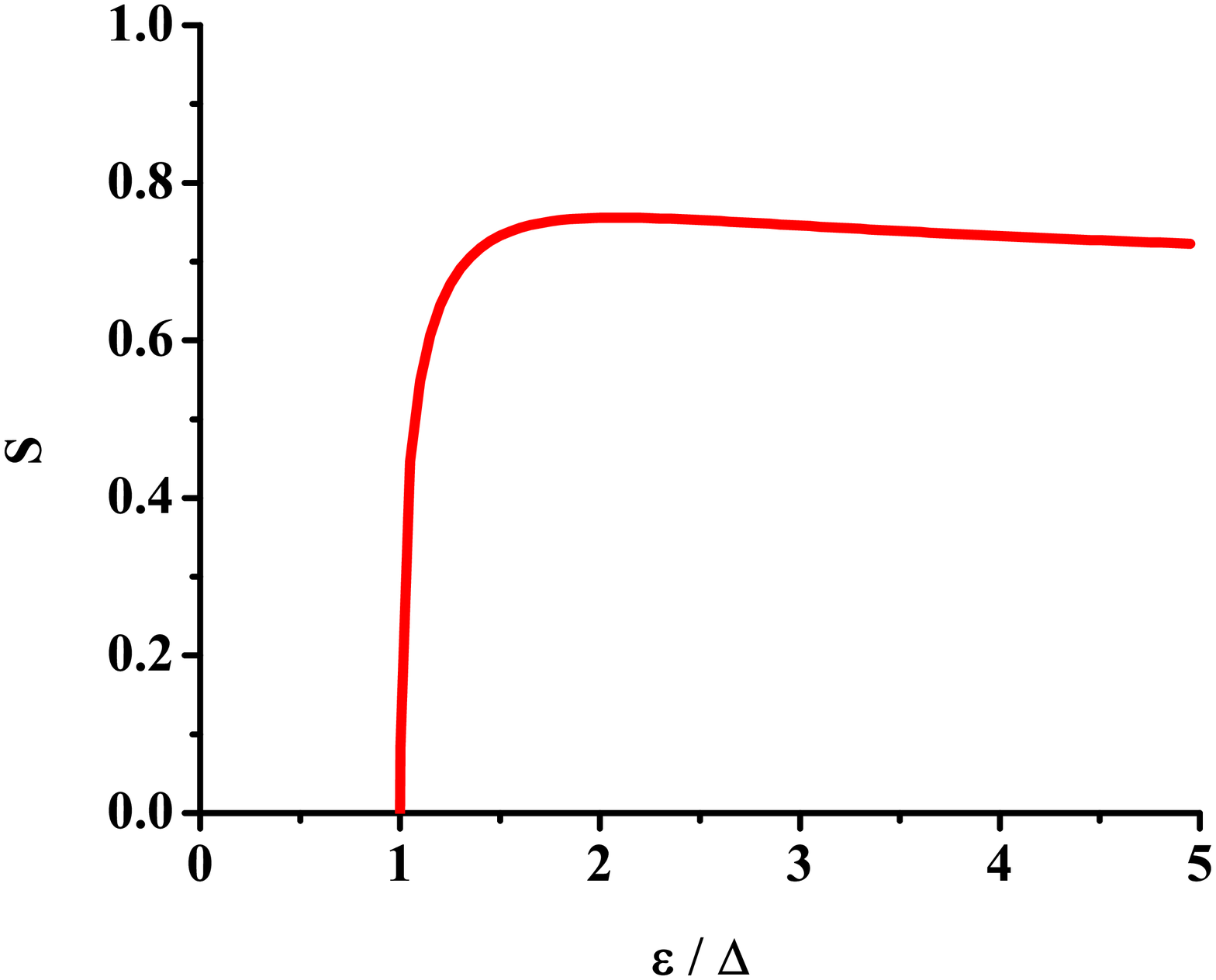}
\caption{ (Color online) The conductance of a normal metal- 3DDM junction vs. excitation energy of carriers normalized by energy of gap of Dirac material (left). The Fano factor of normal metal-3DDM junction vs. excitation energy of carriers normalized by energy of gap of Dirac material (right).}
\label{conductanceM-3DDM}
\end{center}
\end{figure}

The transmission probability, Eq.~(\ref{totalt}) can be used to to calculate the conductance 
and the Fano factor of the junction as,
\begin{equation}
G=G_0\sum_n T_n,
\label{conductance}
\end{equation}
and
\begin{equation}
F=\frac{\sum_n (1-T_n)T_n}{\sum_n T_n},
\end{equation}
where $G_0$ is the conductance quantum $e^2/h$. The Fano factor is the
ratio between the current fluctuations and the average current.
The counter $n$ of the channels in this case in replaced by the wave-vector
$\vec q$. Fig.~\ref{conductanceM-3DDM} shows the result for the conductance and 
the Fano factor as a function of the energy $\eps$ in the 3DMM side. 
As can be seen the behavior is consistend with limiting behavior of
transmission coefficients. For $\eps\gtrsim \Delta$ we have $T\to 1$ so that 
hence $G/G_0\to 1$ per channel, and $F\to 0$. In the other limit $\eps \gg \Delta$,
$T\to 1/4$ and hence $G/G_0\to 1/4$ per channel and $F\to 3/4$. 


\subsection{3DDM sandwiched between two metallic regions}
In order to experimentally measure the quantum transport through 3DDM, it must be connected to
at least two wires from both sides. This corresponds to the situation depicted in 
Fig.~\ref{fig:structure}-b. In this case there will be two interfaces at $z=0$ and
$z=L$ separating the normal metallic region from the 3DDM sandwiched between them.
Similar to the case of the interface between a metal
and 3DDM, here again we can construct the wave function in the left metal corresponding 
to spin $\up$ electron hitting the $z=0$ junction as,
\begin{equation}
\Phi_{m}^{L}=
\left( 
\begin{array}{c}
i\\
0\\
1\\
0\\
\end{array}
\right)
e^{i k_z z} +r_{\uparrow}
\left(
\begin{array}{c}
i\\
0\\
-1\\
0\\
\end{array}
\right)
e^{-i k_z z}
+r_{\downarrow}
\left(
\begin{array}{c}
0\\
-1\\
0\\
i\\
\end{array}
\right)
e^{-i k_z z}.
\end{equation}
Within the 3DDM region, $0\leq z\leq L$, we have,
\begin{equation}
   \psi=c_1 \psi_{D\up}^{+}+c_2 \psi_{D\up}^{-}+c_3\psi_{D\down}^{+}+c_4 \psi_{D\down}^{-}.
\end{equation}
Finally for right metal region we have,
\begin{equation}
\Phi_m^{R}=t_{\uparrow}
\left(
\begin{array}{c}
i\\
0\\
1\\
0\\
\end{array}
\right)
e^{i k_z z}+
t_{\downarrow}
\left(
\begin{array}{c}
0\\
1\\
0\\
i\\
\end{array}
\right)
e^{i k_z z}.
\end{equation}

Using the four-component wave-functions for the normal metallic
regions along with the the following boundary conditions,
\be
\Phi_{m}^{L}\mid_{z=0}=\psi\mid_{z=0},~~~~~~
\Phi_{m}^{R}\mid_{z=L}=\psi\mid_{z=L},\nn
\ee
we obtain amazingly simple result for the transmission amplitudes
$t_\uparrow$ and spin down $t_\downarrow$:
\be
t_{\uparrow}=\frac{1}{\cos(q_zL)-\frac{i \eps}{v_D q_z}\sin(q_zL)},~~~~~~~~~
t_{\downarrow}=0
\label{tuptdown}
\ee
It is interesting to compare the above result with Eq.~\eqref{tdown}. When 
an spin $\up$ electron enters the 3DDM region at a direction not perpendicular to 
the interface, it always has an amplitude $t_\down$ given by Eq.~\eqref{tdown} to
enter the 3DDM. This is due to the fact that in the 3DDM region the motion
of the particle and its spin direction affect each other by strong spin-orbit 
coupling encoded in the Dirac Hamiltonian.
However according to the above equation, when another metallic
region is connected to the right of 3DDM, the final electrons reaching the
right metallic region can only be in spin $\up$ state. 
This can be interpreted as follows: The spin-orbit interaction at each interface
allows for spin-flip transmission. However the spin-flip transmission in the two
interfaces exactly cancel each other and with two interfaces we only get
a net spin-non-flip transmission.

Let us further analyze the transmission through a 3DDM given by Eq.~\eqref{tuptdown}.
This equation implies that for
those modes whose longitudinal wave-vector $q_z$ satisfies $q_zL=n\pi$, the
transmission coefficient is always $1$ irrespective of the energy and the transverse
component. This is a quite natural generalization of the transmission through 
graphene that offers two dimensional example of a Dirac material~\cite{Katsnelson2006}.

\begin{figure}[tb]
\begin{center}
   \vspace{-1.5cm}
\includegraphics[width=5cm]{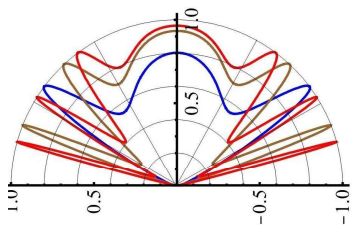}
\caption{(Color online) angular dependence of Eq.~\eqref{angularT.eqn} for $\eps/\Delta=2, 3, 4$ 
corresponding to red, brown and blue curves, respectively. 
The other parameters of junction are $L=1$ $\mu$m, $\Delta=10$ meV and $v_D=10^5$ m/s.}
\label{angularT.fig}
\end{center}
\end{figure}

To discuss another interesting aspect of Eq.~\eqref{tuptdown} let us remember
that $v_Dq=\sqrt{\eps^2-\Delta^2}$ and $q_z=q\cos\theta$ where $\theta$ is the
incidence angle with respect to the direction normal to the interface. When 
the 3DDM becomes gap-less, and the propagation direction is normal to the
interface, the ratio $\eps/(v_Dq)$ in Eq.~\eqref{tuptdown} becomes $1$ and
the transmission amplitude for the incident electron will become,
\be
   t_\up=\exp(-i q L).
\ee
Therefore carriers hitting the gap-less 3DDM region normal to the interface 
always pass through it with probability equal to $1$. This is in some sense
similar to the Klein paradox. One should however bear in mind that the
Klein paradox discusses the tunneling of Dirac electrons form a potential
barrier and under normal incidence of massless Dirac fermions one gets
a transmission probability equal to $1$. But in the present case we find
that massless 3DDM is completely transparent to electrons hitting 
perpendicularly their interface with a normal metal.

Now let us use the definition of $q_z$ to rewrite Eq.~\eqref{tuptdown} as,
\begin{equation}
t_{\uparrow}=\frac{1}{\cos\left( \frac{\sqrt{\eps^2-\Delta^2}\cos\theta}{v_D}L\right) -i\frac{  \eps}{\sqrt{\eps^2-\Delta^2}\cos(\theta)}\sin(\frac{\sqrt{\eps^2-\Delta^2}\cos\theta}{v_D}L)}
\label{angularT.eqn}
\end{equation}
This has been plotted in Fig.~\ref{angularT.fig}.
Typical value of the energy gap $\Delta$, e.g. in a candidate material such as Bismuth 
is of order of $5-15$ meV. The typical value of $v_D$ for the same material is about 
$10^5 m/s$. Therefore when we deal with 3DDM layers with thickness $\sim\mu m$, 
the effect of energy gap is important. When the 3DDM layer become thinner, 
the $L \Delta/v_D$ ratio become smaller and therefore the the transport properties
will not be sensitive to energy of gap and the system is expected to behave
in such a way as if there is no gap.
\begin{figure}[tb]
\begin{center}
\includegraphics[width=6.8cm]{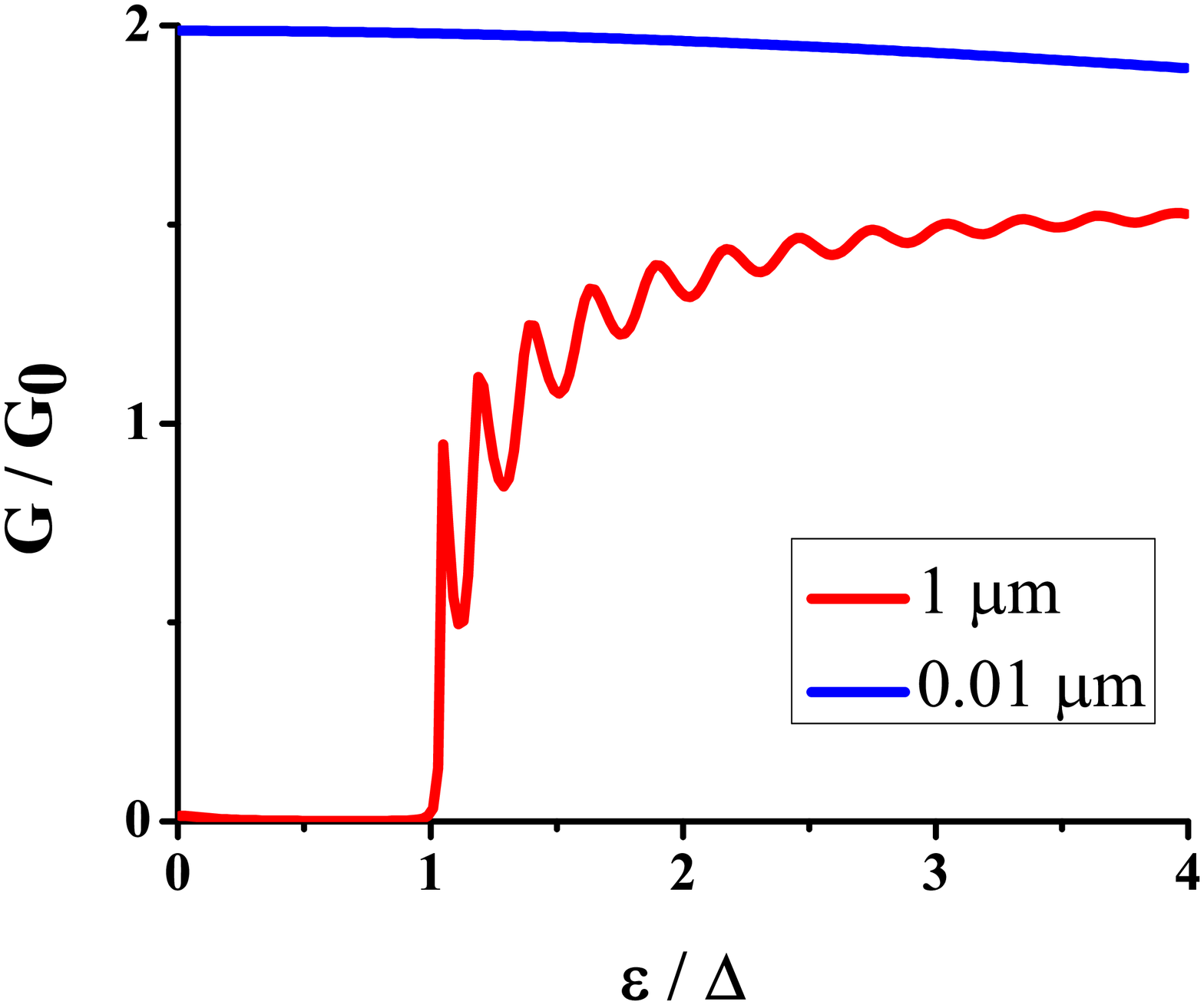}
\includegraphics[width=6.8cm]{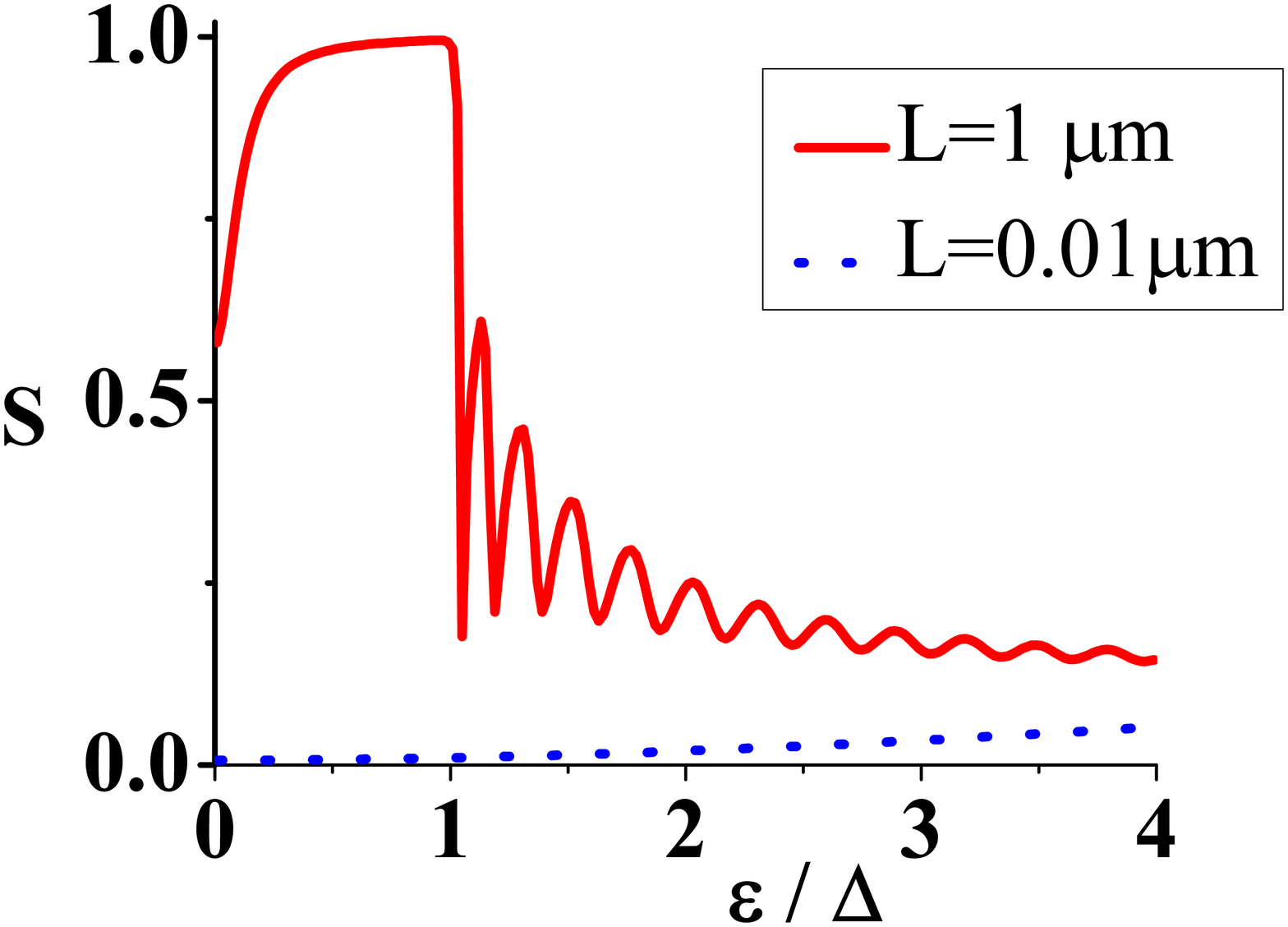}
\caption{(Color online) Conductance (left) and the Fano factor of normal metal-3DDM-normal metal structure.}
\label{gs.fig}
\end{center}
\end{figure}
In Fig.~\ref{gs.fig} we plot the conductance and Fano factor for
two different length scales reflecting this point.

When the transverse direction in the $x$ and $y$ directions confined to lengths 
$W_x$ and $W_y$, respectively, the corresponding transverse wave vectors 
will be quantized. A zero energy electron at $\eps=0$ hitting the 3DDM from 
left will not enter the 3DDM as a propagating mode. Hence the current through 
the 3DDM will be carried as an evanescent mode. The transmission probability
in this case becomes~\cite{Berry1987},
\begin{equation}
t_{\uparrow}=\frac{1}{\cosh\left( \sqrt{\frac{\Delta^2}{v_D^2}+\left( m+\frac{1}{2}\right) ^2\left( \frac{\pi}{W_x}\right) ^2+\left( n+\frac{1}{2}\right) ^2\left( \frac{\pi}{W_y}\right) ^2}\right) },
\end{equation} 
where $m, n$ label the quantized transverse channels. 
This can be considered as the three dimensional and gapped generalization of a
similar result found earlier for graphene~\cite{Beenakker2007}.

\section{Summary and outlook}
In this work we studied the transport through a three dimensional Dirac material 
whose low-energy electronic states are described by massive Dirac Hamiltonian.
This theory is specified by a gap parameter $\Delta$ and a velocity scale $v_D$
that is usually two or three orders of magnitude smaller than the light velocity.
We constructed an artificial four component wave function in the normal metallic
region in such a way that the current resulting from this four-component wave 
function gives precisely the current arising from the corresponding Schr\"odinger
equation. Then the matching condition between a normal metal and the 3DDM can 
be applied that guarantees

Considering a single interface between a normal metal and a 3DDM we found
that electrons hitting the interface at an angle can enter the 3DDM as
spin-flipped due to spin-orbit coupling in the 3DDM. For a 3DDM of finite
length sandwiched between two normal metallic region we found that 
the spin-flip transmission at the
second interface exactly cancels the one at the opposite interface. 
We found further that at normal incidence when the gap parameter is
zero, the 3DDM becomes completely transparent. This is in some sense
similar to the Klein tunneling. We also found that electrons 
transmitted at the energy corresponding to the bottom of the Dirac conduction
band (or holes corresponding to the top of the Dirac valence band) pass 
through the 3DDM with probability $1$. The 3DDM can also provide transmission
through evanescent modes when the incident particle energy corresponds to 
the mid-gap states of the 3DDM.

\section{Acknowledgements}
This research was completed while the visit of SAJ to the university of Duisburg
supported the Alexander von Humboldt foundation.

\section*{References}

\bibliography{Quantum}

\end{document}